\documentclass[aps,prl,preprint,floatfix,superscriptaddress,showpacs]{revtex4}

\usepackage{graphicx}
\usepackage{dcolumn}
\usepackage{ulem}
\usepackage{soul}

\begin{document}
\newcommand{\el}{\mbox{${\rm e^{-}}$ }}
\newcommand{\ps}{\mbox{${\rm e^{+}}$ }}

\title{The cosmic-ray electron flux measured by the PAMELA experiment
  between 1 and 625 GeV}

\author{O. Adriani}
\affiliation{University of Florence, Department of Physics,  
 I-50019 Sesto Fiorentino, Florence, Italy}
\affiliation{INFN, Sezione di Florence,  
 I-50019 Sesto Fiorentino, Florence, Italy}
\author{G. C. Barbarino}
\affiliation{University of Naples ``Federico II'', Department of
Physics, I-80126 Naples, Italy}
\affiliation{INFN, Sezione di Naples,  I-80126 Naples, Italy}
\author{G. A. Bazilevskaya}
\affiliation{Lebedev Physical Institute, RU-119991
Moscow, Russia}
\author{R. Bellotti}
\affiliation{University of Bari, Department of Physics, I-70126 Bari, Italy}
\affiliation{INFN, Sezione di Bari, I-70126 Bari, Italy}
\author{M. Boezio}
\affiliation{INFN, Sezione di Trieste, I-34149
Trieste, Italy}
\author{E. A. Bogomolov}
\affiliation{Ioffe Physical Technical Institute,  RU-194021 St. 
Petersburg, Russia}
\author{M. Bongi}
\affiliation{INFN, Sezione di Florence,  
 I-50019 Sesto Fiorentino, Florence, Italy}
\author{V. Bonvicini}
\affiliation{INFN, Sezione di Trieste,  I-34149
Trieste, Italy}
\author{S. Borisov}
\affiliation{INFN, Sezione di Rome ``Tor Vergata'', I-00133 Rome, Italy}
\affiliation{University of Rome ``Tor Vergata'', Department of
Physics,  I-00133 Rome, Italy} 
\affiliation{National Research Nuclear University ``MEPHI'',  RU-115409
Moscow, Russia}  
\author{S. Bottai}
\affiliation{INFN, Sezione di Florence,  
 I-50019 Sesto Fiorentino, Florence, Italy}
\author{A. Bruno}
\affiliation{University of Bari, Department of Physics, I-70126 Bari,
Italy} 
\affiliation{INFN, Sezione di Bari, I-70126 Bari, Italy}
\author{F. Cafagna}
\affiliation{INFN, Sezione di Bari, I-70126 Bari, Italy}
\author{D. Campana}
\affiliation{INFN, Sezione di Naples,  I-80126 Naples, Italy}
\author{R. Carbone}
\affiliation{INFN, Sezione di Naples,  I-80126 Naples, Italy}
\affiliation{University of Rome ``Tor Vergata'', Department of
Physics,  I-00133 Rome, Italy} 
\author{P. Carlson}
\affiliation{KTH, Department of Physics, and the Oskar Klein Centre for
Cosmoparticle Physics, AlbaNova University Centre, SE-10691 Stockholm,
Sweden}
\author{M. Casolino}
\affiliation{INFN, Sezione di Rome ``Tor Vergata'', I-00133 Rome, Italy}
\author{G. Castellini}
\affiliation{ IFAC,  I-50019 Sesto Fiorentino,
Florence, Italy}
\author{L. Consiglio}
\affiliation{INFN, Sezione di Naples,  I-80126 Naples, Italy}
\author{M. P. De Pascale}
\affiliation{INFN, Sezione di Rome ``Tor Vergata'', I-00133 Rome, Italy}
\affiliation{University of Rome ``Tor Vergata'', Department of
Physics,  I-00133 Rome, Italy} 
\author{C. De Santis}
\affiliation{INFN, Sezione di Rome ``Tor Vergata'', I-00133 Rome, Italy}
\author{N. De Simone}
\affiliation{INFN, Sezione di Rome ``Tor Vergata'', I-00133 Rome, Italy}
\affiliation{University of Rome ``Tor Vergata'', Department of
Physics,  I-00133 Rome, Italy} 
\author{V. Di Felice}
\affiliation{INFN, Sezione di Rome ``Tor Vergata'', I-00133 Rome, Italy}
\author{A. M. Galper}
\affiliation{National Research Nuclear University ``MEPHI'',  RU-115409
Moscow, Russia}  
\author{W. Gillard}
\affiliation{KTH, Department of Physics, and the Oskar Klein Centre for
Cosmoparticle Physics, AlbaNova University Centre, SE-10691 Stockholm,
Sweden}
\author{L. Grishantseva}
\affiliation{National Research Nuclear University ``MEPHI'',  RU-115409
Moscow, Russia}  
\author{G. Jerse}
\affiliation{INFN, Sezione di Trieste, I-34149
Trieste, Italy}
\affiliation{University of Trieste, Department of Physics, 
I-34147 Trieste, Italy}
\author{A. V. Karelin}
\affiliation{National Research Nuclear University ``MEPHI'',  RU-115409
Moscow, Russia}
\author{S. V. Koldashov}
\affiliation{National Research Nuclear University ``MEPHI'',  RU-115409
Moscow, Russia}  
\author{S. Y. Krutkov}
\affiliation{Ioffe Physical Technical Institute,  RU-194021 St. 
Petersburg, Russia}
\author{A. N. Kvashnin}
\affiliation{Lebedev Physical Institute, RU-119991
Moscow, Russia}
\author{A. Leonov}
\affiliation{National Research Nuclear University ``MEPHI'',  RU-115409
Moscow, Russia}  
\author{V. Malakhov}
\affiliation{National Research Nuclear University ``MEPHI'',  RU-115409
Moscow, Russia}  
\author{V. Malvezzi}
\affiliation{INFN, Sezione di Rome ``Tor Vergata'', I-00133 Rome, Italy}
\author{L. Marcelli}
\affiliation{INFN, Sezione di Rome ``Tor Vergata'', I-00133 Rome, Italy}
\author{A. G. Mayorov}
\affiliation{National Research Nuclear University ``MEPHI'',  RU-115409
Moscow, Russia}
\author{W. Menn}
\affiliation{Universit\"{a}t Siegen, Department of Physics,
D-57068 Siegen, Germany}
\author{V. V. Mikhailov}
\affiliation{National Research Nuclear University ``MEPHI'',  RU-115409
Moscow, Russia}  
\author{E. Mocchiutti}
\affiliation{INFN, Sezione di Trieste,  I-34149
Trieste, Italy}
\author{A. Monaco}
\affiliation{University of Bari, Department of Physics, I-70126 Bari, Italy}
\affiliation{INFN, Sezione di Bari, I-70126 Bari, Italy}
\author{N. Mori}
\affiliation{University of Florence, Department of Physics,  
 I-50019 Sesto Fiorentino, Florence, Italy}
\affiliation{INFN, Sezione di Florence,  
 I-50019 Sesto Fiorentino, Florence, Italy}
\author{N. Nikonov}
\affiliation{Ioffe Physical Technical Institute,  RU-194021 St. 
Petersburg, Russia}
\affiliation{INFN, Sezione di Rome ``Tor Vergata'', I-00133 Rome, Italy}
\affiliation{University of Rome ``Tor Vergata'', Department of
Physics,  I-00133 Rome, Italy} 
\author{G. Osteria}
\affiliation{INFN, Sezione di Naples,  I-80126 Naples, Italy}
\author{F. Palma}
\affiliation{INFN, Sezione di Rome ``Tor Vergata'', I-00133 Rome, Italy}
\affiliation{University of Rome ``Tor Vergata'', Department of
Physics,  I-00133 Rome, Italy} 
\author{P. Papini}
\affiliation{INFN, Sezione di Florence,  
 I-50019 Sesto Fiorentino, Florence, Italy}
\author{M. Pearce}
\affiliation{KTH, Department of Physics, and the Oskar Klein Centre for
Cosmoparticle Physics, AlbaNova University Centre, SE-10691 Stockholm,
Sweden}
\author{P. Picozza}
\affiliation{INFN, Sezione di Rome ``Tor Vergata'', I-00133 Rome, Italy}
\affiliation{University of Rome ``Tor Vergata'', Department of
Physics,  I-00133 Rome, Italy} 
\author{C. Pizzolotto}
\affiliation{INFN, Sezione di Trieste, I-34149
Trieste, Italy}
\author{M. Ricci}
\affiliation{INFN, Laboratori Nazionali di Frascati, Via Enrico Fermi 40,
I-00044 Frascati, Italy}
\author{S. B. Ricciarini}
\affiliation{INFN, Sezione di Florence, 
 I-50019 Sesto Fiorentino, Florence, Italy}
\author{L. Rossetto}
\affiliation{KTH, Department of Physics, and the Oskar Klein Centre for
Cosmoparticle Physics, AlbaNova University Centre, SE-10691 Stockholm,
Sweden}
\author{R. Sarkar}
\affiliation{INFN, Sezione di Trieste, I-34149
Trieste, Italy}
\author{M. Simon}
\affiliation{Universit\"{a}t Siegen, Department of Physics,
D-57068 Siegen, Germany}
\author{R. Sparvoli}
\affiliation{INFN, Sezione di Rome ``Tor Vergata'', I-00133 Rome, Italy}
\affiliation{University of Rome ``Tor Vergata'', Department of
Physics,  I-00133 Rome, Italy} 
\author{P. Spillantini}
\affiliation{University of Florence, Department of Physics,  
 I-50019 Sesto Fiorentino, Florence, Italy}
\affiliation{INFN, Sezione di Florence,  
 I-50019 Sesto Fiorentino, Florence, Italy}
\author{S. J. Stochaj}
\affiliation{New Mexico State University, Las Cruces, NM 88003, USA}
\author{J. C. Stockton}
\affiliation{New Mexico State University, Las Cruces, NM 88003, USA}
\author{Y. I. Stozhkov}
\affiliation{Lebedev Physical Institute, RU-119991
Moscow, Russia}
\author{A. Vacchi}
\affiliation{INFN, Sezione di Trieste,  I-34149
Trieste, Italy}
\author{E. Vannuccini}
\affiliation{INFN, Sezione di Florence, 
 I-50019 Sesto Fiorentino, Florence, Italy}
\author{G. Vasilyev}
\affiliation{Ioffe Physical Technical Institute, RU-194021 St. 
Petersburg, Russia}
\author{S. A. Voronov}
\affiliation{National Research Nuclear University ``MEPHI'',  RU-115409
Moscow, Russia}
\author{J. Wu}
\altaffiliation[On leave from ]{School of Mathematics and Physics,
China University of Geosciences, CN-430074 Wuhan, China}
\affiliation{KTH, Department of Physics, and the Oskar Klein Centre for
Cosmoparticle Physics, AlbaNova University Centre, SE-10691 Stockholm,
Sweden}
\author{Y. T. Yurkin}
\affiliation{National Research Nuclear University ``MEPHI'',  RU-115409
Moscow, Russia}  
\author{G. Zampa}
\affiliation{INFN, Sezione di Trieste,  I-34149
Trieste, Italy}
\author{N. Zampa}
\affiliation{INFN, Sezione di Trieste,  I-34149
Trieste, Italy}
\author{V. G. Zverev}
\affiliation{National Research Nuclear University ``MEPHI'',  RU-115409
Moscow, Russia}  

\date{\today}

\begin{abstract}

Precision measurements of the electron component in the cosmic radiation
provide important information about the origin and propagation of
cosmic rays in the Galaxy. 
Here we present new results regarding 
negatively charged electrons between 1 and 625 GeV performed by the 
satellite-borne experiment PAMELA. This is the first time that cosmic-ray
\el have been identified above 50~GeV.  
The electron spectrum can be described
with a single power law energy dependence with spectral index 
$-3.18 \pm 0.05$ 
above the energy region influenced by the solar wind ($> 30$~GeV). No
significant spectral features are observed and the data can be
interpreted in terms of conventional diffusive propagation 
models. However, the data
are also consistent with models including new cosmic-ray sources 
that could explain the rise in the positron fraction.

\end{abstract}

\pacs{98.70.Sa, 96.50.sb, 95.35.+d}

\maketitle


Cosmic-ray electrons are a small but important component of the
cosmic 
radiation. They provide information regarding the origin and
propagation of cosmic rays in the Galaxy that is not accessible from
the study of the cosmic-ray nuclear components due to their differing
energy-loss processes.
Cosmic-ray
electrons and positrons are 
produced as secondaries by the interactions  between
cosmic-ray nuclei and the interstellar matter. However, since the
observed positron fraction (\mbox{$\phi$(e$^+$) / ($\phi$(e$^+$) +
  $\phi$(e$^-$))}, where $\phi$ is the flux,
is of the order of ten percent and less
above a few GeV \cite{des64,fan69,adr09b}, a majority
of electrons must be of primary origin. 

Due to 
their low mass and the intergalactic magnetic
field, cosmic-ray electrons undergo severe energy losses 
during their propagation in the Galaxy. Therefore, 
it can be expected that a significant fraction of high
energy ($> 10$~GeV) electrons and positrons are produced 
in the solar neighborhood ($\sim 1$ kpc) \cite{del09}
with the majority of the primary electron component probably
originating from a small number 
of sources, which may induce features in the spectral shape of the
electron energy spectrum \cite{nis80,del10}. 
Spectral features may also arise from the contribution of 
more exotic sources such
as dark
matter particles, e.g.~\cite{cir08}, or other 
astrophysical objects
such as pulsars, e.g.~\cite{ato95}. Both 
were invoked to explain the
positron fraction measured by PAMELA~\cite{adr09b} and 
are expected to contribute to the cosmic
radiation with roughly equal numbers of electrons and positrons. 

Measuring the energy spectrum of cosmic-ray electrons 
involves the 
difficult identification of this rare component and determination of
detector efficiencies and particle 
energies. Therefore, it is not a surprise that results, gathered
mostly by balloon-borne experiments in the past decades, differ
beyond quoted errors.
Another point that has to be
highlighted is that there are no measurements of the high energy
(above $\sim 50$~GeV) negatively-charged electron flux. 

The results presented here are based on the data-set collected by
the PAMELA satellite-borne experiment between July 2006
and January 2010. From over  
$2 \times 10^9$ triggered events, accumulated during a total
acquisition time of 
approximately 1200~days,  
377,614 electrons were selected 
in the energy interval  
1 - 625 GeV, the largest energy range covered in any cosmic ray \el
experiment hitherto. 
Further details on the PAMELA apparatus, orbit and data acquisition
can be found in~~\cite{pic07,boe09,adr10b}. 

A sample of negatively-charged particles was selected using the 
time-of-flight and spectrometer data. This consisted 
mostly of
electrons with a few percent contamination of cosmic-ray antiprotons. 
At higher
rigidities ``spillover'' protons, reconstructed with an
incorrect 
sign of curvature either due to the finite
spectrometer resolution or scattering in the spectrometer planes,
represented the largest source of contamination estimated to 
increase from a 
few percent at $\sim 100$~GV/c to about ten times the
electron signal around  
500~GV/c. All these contamination components were reduced to a negligible
amount by requiring an electromagnetic-like
interaction pattern in the 16 radiation length deep
silicon-tungsten calorimeter~\cite{boe06,boe09}. 
Electrons were selected up to $\sim 600$~GV/c.
Above this 
rigidity the sign-of-curvature of tracks could not be reliably
resolved due to statistical and systematic uncertainties.

The most important contributions to the 
discrepancies between the various electron measurements are
instrumental effects such as selection efficiencies and energy
determination. To reduce the systematic uncertainties 
the selection efficiencies were derived from flight data,
cross-checking the results with those obtained using simulations of
the apparatus based both 
on the GEANT3~\cite{bru94} and GEANT4~\cite{ago03} packages. The
validity of the 
simulations was confirmed by comparisons with test-beam and flight
data. The simulations were also used in PAMELA results concerning 
antiprotons, protons and
helium nuclei \cite{adr10b,adr10c}.
The total systematic uncertainty on the flux was found to 
increase from about 4\% at 1 GV/c to about 7\% at 600 GV/c. This 
uncertainty was obtained quadratically summing the various systematic
errors considered: acceptance, efficiency
estimation and spectrum unfolding.
The energy-binned electron fluxes 
are given in Table~\ref{t:elet}. 
\begingroup
\squeezetable
\begin{table}[H] 
\caption{Summary of electron 
results. The first and
second errors represent the statistical and 
systematic uncertainties, respectively. The mean kinetic energy has
been obtained following the procedure described
in~\cite{laf95}. \label{t:elet}}    
\begin{ruledtabular}
\begin{tabular}{cccc}
Rigidity & Mean Kinetic & Observed & Flux \\ 
at the & Energy at & number of & at top of \\  
spectrometer & top of & events & payload \\
GV/c & payload GeV & 
 &  (particles/(m$^{2}$ sr s
GeV)) \\ \hline
    1.04 -   1.19 &   1.11 & 27930 & $  31.2 \pm    0.2 \pm    1.3$ \\
    1.19 -   1.37 &   1.28 & 30361 & $  26.7 \pm    0.2 \pm    1.1$ \\
    1.37 -   1.57 &   1.47 & 32973 & $  23.0 \pm    0.1 \pm    1.0$ \\
    1.57 -   1.80 &   1.68 & 33787 & $  18.8 \pm    0.1 \pm    0.8$ \\
    1.80 -   2.07 &   1.93 & 33613 & $ 15.03 \pm   0.08 \pm    0.6$ \\
    2.07 -   2.38 &   2.22 & 32854 & $ 11.94 \pm   0.07 \pm    0.5$ \\
    2.38 -   2.73 &   2.55 & 30118 & $  8.97 \pm   0.05 \pm    0.4$ \\
    2.73 -   3.13 &   2.92 & 27234 & $  6.67 \pm   0.04 \pm    0.3$ \\
    3.13 -   3.60 &   3.36 & 23607 & $  4.76 \pm   0.03 \pm    0.2$ \\
    3.60 -   4.13 &   3.85 & 20440 & $  3.40 \pm   0.02 \pm    0.1$ \\
    4.13 -   4.74 &   4.42 & 16817 & $  2.30 \pm   0.02 \pm    0.1$ \\
     4.7 -    5.4 &    5.1 & 13812 & $  1.56 \pm   0.01 \pm   0.07$ \\
     5.4 -    6.3 &    5.8 & 11428 & $  1.06 \pm   0.01 \pm   0.05$ \\
     6.3 -    7.2 &    6.7 &  9410 & $(   7.2 \pm   0.07 \pm    0.3) \times 10^{-1} $ \\
     7.2 -    8.2 &    7.7 &  7374 & $(   4.7 \pm   0.05 \pm    0.2) \times 10^{-1} $ \\
     8.2 -    9.5 &    8.8 &  5851 & $(   3.1 \pm   0.04 \pm    0.1) \times 10^{-1} $ \\
     9.5 -   10.9 &   10.1 &  4441 & $(  1.91 \pm   0.03 \pm   0.08) \times 10^{-1} $ \\
    10.9 -   12.5 &   11.6 &  3583 & $(  1.26 \pm   0.02 \pm   0.05) \times 10^{-1} $ \\
    12.5 -   14.3 &   13.4 &  2767 & $(   7.9 \pm    0.2 \pm    0.3) \times 10^{-2} $ \\
    14.3 -   16.4 &   15.3 &  2266 & $(   5.2 \pm    0.1 \pm    0.2) \times 10^{-2} $ \\
    16.4 -   18.9 &   17.6 &  1798 & $(  3.26 \pm   0.08 \pm    0.1) \times 10^{-2} $ \\
    18.9 -   21.7 &   20.2 &  1392 & $(  2.08 \pm   0.06 \pm   0.09) \times 10^{-2} $ \\
    21.7 -   24.9 &   23.2 &   972 & $(  1.26 \pm   0.04 \pm   0.05) \times 10^{-2} $ \\
    24.9 -   28.6 &   26.6 &   778 & $(   8.9 \pm    0.3 \pm    0.4) \times 10^{-3} $ \\
    28.6 -   32.8 &   30.6 &   518 & $(   5.2 \pm    0.2 \pm    0.2) \times 10^{-3} $ \\
    32.8 -   37.7 &   35.1 &   422 & $(   3.7 \pm    0.2 \pm    0.2) \times 10^{-3} $ \\
    37.7 -   43.3 &   40.3 &   276 & $(   2.2 \pm    0.1 \pm   0.09) \times 10^{-3} $ \\
    43.3 -   49.7 &   46.3 &   211 & $(   1.4 \pm    0.1 \pm   0.06) \times 10^{-3} $ \\
    49.7 -   57.0 &   53.2 &   172 & $(  1.04 \pm   0.08 \pm   0.04) \times 10^{-3} $ \\
    57.0 -   65.5 &   61.0 &   104 & $(   5.5 \pm    0.5 \pm    0.2) \times 10^{-4} $ \\
    65.5 -   75.2 &   70.1 &    87 & $(   4.1 \pm    0.4 \pm    0.2) \times 10^{-4} $ \\
    75.2 -   86.3 &   80.5 &    52 & $(   2.1 \pm    0.3 \pm   0.09) \times 10^{-4} $ \\
    86.3 -   99.1 &   92.4 &    42 & $(   1.5 \pm    0.2 \pm   0.07) \times 10^{-4} $ \\
    99.1 -  119.1 &  108.5 &    41 & $(   10. \pm     2. \pm    0.4) \times 10^{-5} $ \\
   119.1 -  143.2 &  130.4 &    33 & $(    7. \pm     1. \pm    0.3) \times 10^{-5} $ \\
   143.2 -  188.8 &  163.8 &    25 & $(   2.7 \pm    0.5 \pm    0.1) \times 10^{-5} $ \\
   188.8 -  260.7 &  220.7 &    14 & $(   9.6^{+3.0}_{-2.5} \pm    0.5) \times 10^{-6} $ \\
   260.7 -  394.5 &  317.9 &     7 & $(   2.8^{+1.3}_{-1.1} \pm    0.1) \times 10^{-6} $ \\
   394.5 -  625.3 &  491.4 &     3 & $(   9^{+7}_{-6} \pm
   1) \times 10^{-7} $ \\
\end{tabular}
\end{ruledtabular}
\end{table}
\endgroup
These results were obtained using the rigidity
measured by the magnetic spectrometer and unfolding 
the resulting energy spectrum to the top of the payload using a
Bayesian approach, as described in~\cite{dag95}. This unfolding was
particularly important for electrons (and positrons) due to their
non-negligible energy losses, primarily due to bremsstrahlung while
traversing the pressurised container and parts of the apparatus prior
to the tracking system (equivalent to about 0.1 radiation lengths).

Since the PAMELA calorimeter was also designed to
precisely sample the total 
energy deposited by electromagnetic showers~\cite{boe02}, this
information was used to derive the
energy of the impinging electron. 
Containment requirements (at least half Moliere radii from the silicon 
detector borders for each calorimeter layer) were applied to the projected 
track in the calorimeter. This resulted in a good Gaussian energy 
resolution, varying from $\simeq 8\%$ at 10 GeV to $\simeq 3\%$ above 
100~GeV, but also in a decrease in statistics of $\simeq 50\%$.
Hence, it was possible to obtain 
an estimation of the energy of cosmic-ray electrons
that was systematically independent of the rigidity
measurement, 
The bremsstrahlung photons, produced by electrons while crossing the
top part of the payload, 
converted into electromagnetic
showers in the calorimeter, thus  
allowing the total energy of the incoming electron to be
estimated. Therefore,  
the calorimetry measurement provided 
a cross-check of the
energy spectrum derived from the tracking system information.

\begin{figure}[h]
\includegraphics[width=25pc]{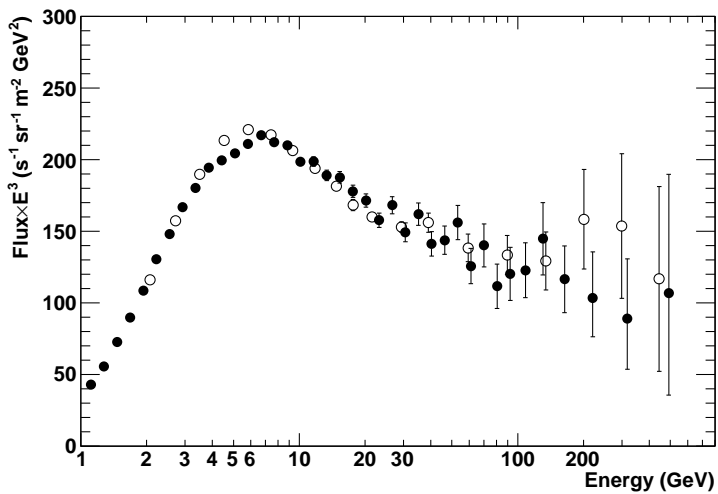}\hspace{2pc}
\caption{The negatively-charged electron spectrum measured by PAMELA
  with two independent approaches:
  energy derived from the rigidity (full circles); energy derived from
  the calorimeter information (open circles). 
  The error bars are statistical only.
\label{fluxtrcal}}   
\end{figure}

Figure~\ref{fluxtrcal} shows the electron energy spectra obtained using
the calorimeter and the tracking information. The sign of the
curvature in the magnetic spectrometer was used to select negative
particles also for the calorimeter case, thus making a
consistent comparison possible. 
The two sets of measurements are in good agreement  
considering the uncertainty of the
reconstruction and unfolding procedures. 
The results discussed in this work are based on the magnetic 
spectrometer rigidity that provided a larger statistical sample and a 
better energy resolution in the most statistically significant energy 
region.

\begin{figure}[h]
\includegraphics[width=25pc]{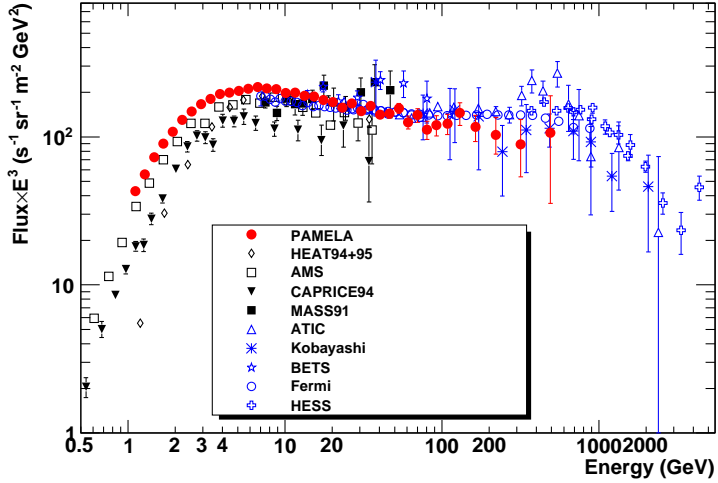}\hspace{2pc}
\caption{The electron energy spectrum 
obtained in 
this work compared with modern 
measurements:  
CAPRICE94~\protect\cite{boe00}, 
HEAT~\protect\cite{duv01}, 
AMS~\protect\cite{alc00}, MASS91~\protect\cite{gri02},
Kobayashi~\protect\cite{kob99}, BETS~\protect\cite{tor01},
ATIC~\protect\cite{cha08}, 
HESS~\protect\cite{aha08}, Fermi~\protect\cite{ack10}. 
Note that the data points from
\protect\cite{kob99,tor01,cha08,aha08,ack10}, indicated with blue
symbols, and the highest data  
point from HEAT~\protect\cite{duv01} are for the electron and positron
sum.
\label{flux1}}   
\end{figure}

Figure~\ref{flux1} shows the electron energy spectrum measured 
by PAMELA 
along with other recent experimental
data~\cite{boe00,alc00,duv01,gri02,kob99,tor01,cha08,aha08,ack10}.  
The data from
\protect\cite{kob99,tor01,cha08,aha08,ack10} and the highest data
point from HEAT~\protect\cite{duv01} refer to the sum of electron and
positron fluxes.
Considering statistical and systematic uncertainties, no significant
disagreements are found between PAMELA and the recent 
ATIC~\cite{cha08} and Fermi~\cite{ack10} data, even considering an
additional positron component in these measurements of order a few
percent (see~\cite{adr10a}). However, the PAMELA \el spectrum appears 
softer than the (\el + \ps)
spectra presented by ATIC and Fermi. This difference is within the
systematic uncertainties between the various measurements, but it is
also consistent with a growing positron component with energy.
An analysis of the PAMELA positron energy
spectrum (up to $\sim 300$~GeV) 
will be presented in a future publication.
The differences with previous 
magnetic-spectrometer measurements \cite{boe00,duv01,alc00,gri02} are
larger 
and probably due to uncertainties in the energy
and efficiencies determination of the various experiments. Below
10~GeV, discrepancies can be  
partially explained by the effect of solar modulation 
for the various data taking periods. 

\begin{figure}[h]
\includegraphics[width=25pc]{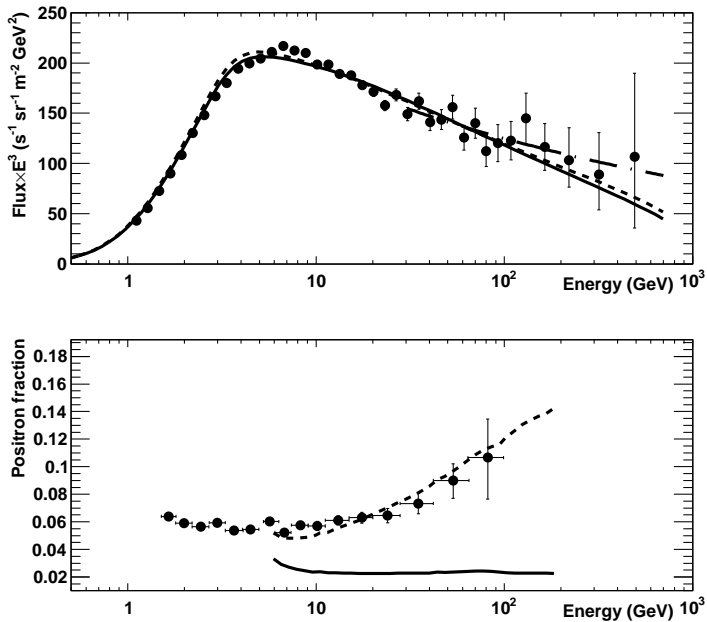}\hspace{2pc}
\caption{Top: the \el spectrum
obtained in 
this work compared with
theoretical calculations. The solid line shows the standard GALPROP
calculation~\protect\cite{str98} 
 for a diffusion reacceleration
model; the long-dashed line is a single power-law fit to the data above
30~GeV; the short-dashed line is a GALPROP calculation including a 
component from additional cosmic-ray electron sources. Bottom: the
PAMELA positron 
fraction~\protect\cite{adr10a} compared with the previous GALPROP
calculations with no (solid line) and with additional
\el and \ps components (short-dashed line). The error bars are
statistical only and these were the only errors considered by the
fitting procedures.    
\label{flux2}}   
\end{figure}

Figure~\ref{flux2} top shows
the PAMELA \el spectrum compared with a theoretical calculation
(solid line) based on the GALPROP code~\cite{str98} and with a single
power-law fit (long-dashed line) to the data above 30~GeV (above the
influence of solar modulation). The single power-law fit represents 
well the data ($\chi^{2} / {\rm ndf} = 8.7 / 13$) with a resulting spectral
index of $-3.18 \pm 0.05$. This is
incompatible (about 6 standard deviation discrepancy even
considering systematic errors)
with the soft \el spectrum~\cite{del09} required
to explain the PAMELA 
positron fraction measurement within a standard model of cosmic-ray
propagation. 
The GALPROP calculation was performed using a spatial Kolmogorov
diffusion with spectral  
index $\delta = 0.34$ and diffusive reacceleration characterized 
by an Alfv\'{e}n speed $v_{A} = 36$~km/s and halo height
of 4 kpc (parameters
from~\cite{ptu06}). The injection
\el spectrum (spectral index: -2.66) was obtained from a
best fit of the propagated spectrum to PAMELA results, which was  
normalized to the data at $\sim 70$~GeV and calculated for
solar minimum, 
using the force field approximation \cite{gle68} ($\Phi = 600$~MV). For 
secondary \el production during propagation we used primary proton
and helium spectra that reproduced PAMELA
measurements~\cite{adr10c}. This 
GALPROP calculation reproduces fairly well the results above 10~GeV
($\chi^{2} / {\rm ndf} = 35 / 26$), however 
differences between the 
measured and predicted spectral shapes can be noticed.  
This may indicate that changes in the propagation model  
or additional sources of cosmic-ray electrons are needed. 
The GALPROP calculation is commonly used
assuming a continuous distribution of sources in the Galaxy.
However, due to the
significant energy losses this
does not seem plausible for primary high energy electrons~\footnote{On
  the contrary this is valid for secondary \el and \ps since the
  cosmic-ray protons and nuclei interact with the ambient gas fairly 
  uniformly in the interstellar medium.}, since this
assumption should only hold for a relatively close
neighborhood. Furthermore, as pointed out in \cite{sha09}, SNRs are
concentrated 
in the spiral arms of the Galaxy, therefore one should consider an
inhomogeneous source distribution. 

One important point concerning sources of  
primary \ps invoked to explain the 
positron fraction measurement \cite{adr09b}, is that they should contribute 
to both the \ps and \el components in
about equal amount. 
It is therefore reasonable to investigate
 if the PAMELA
\el data  can accommodate 
an additional component consistent with the positron
fraction~\cite{adr10a}. 
Hence, we repeated the 
previous GALPROP calculation including an \el
component resulting from new sources for which the only assumption was
that they injected \el and \ps in the interstellar medium with a power
law energy spectrum. The best fit to
the data indicated that a model (short-dashed
line in Figure~\ref{flux2} top) with three components: 
two primary electron components with different
injection spectra ($ 2.69 \pm 0.04$ and $2.1 \pm 0.4$) and secondary
electrons, provided a better agreement
to PAMELA data ($\chi^{2} / {\rm ndf} = 30.9 / 27$)
than the standard, two component GALPROP
calculation (solid line). Furthermore, assuming that the new primary
component, 
which dominated the high energy region with a harder spectrum, 
identically contributed to the positron component we were
able to reproduce PAMELA positron fraction~\cite{adr10a} above
5~GeV. Figure~\ref{flux2} bottom shows this positron fraction compared
to the GALPROP predictions with no (solid line) and with additional
\el and \ps components (short-dashed line). 

We have measured the \el energy spectrum 
over the 
broadest energy range ever achieved and with no atmospheric
overburden. Our results are not 
inconsistent with the standard model of
cosmic ray acceleration and propagation in the Galaxy.
However, there is some tension between the data and the prediction
that points to needed refinements of the propagation models and might
require 
additional sources of cosmic rays. 

\begin{acknowledgments}
We acknowledge support from The Italian Space Agency 
(ASI), Deutsches Zentrum f\"{u}r Luft- und Raumfahrt (DLR), The
Swedish National Space  
Board, The Swedish Research Council, The Russian Space Agency
(Roscosmos) and The
Russian Foundation for Basic Research. R. S. wishes to thank the TRIL
program of the International Center of Theoretical Physics, Trieste,
Italy that partly sponsored his activity.
\end{acknowledgments}

\bibliography{pamela_elflux}

\begin{thebibliography}{32}
\expandafter\ifx\csname natexlab\endcsname\relax\def\natexlab#1{#1}\fi
\expandafter\ifx\csname bibnamefont\endcsname\relax
  \def\bibnamefont#1{#1}\fi
\expandafter\ifx\csname bibfnamefont\endcsname\relax
  \def\bibfnamefont#1{#1}\fi
\expandafter\ifx\csname citenamefont\endcsname\relax
  \def\citenamefont#1{#1}\fi
\expandafter\ifx\csname url\endcsname\relax
  \def\url#1{\texttt{#1}}\fi
\expandafter\ifx\csname urlprefix\endcsname\relax\def\urlprefix{URL }\fi
\providecommand{\bibinfo}[2]{#2}
\providecommand{\eprint}[2][]{\url{#2}}

\bibitem[{\citenamefont{Shong et~al.}(1964)\citenamefont{Shong, Hildebrand, and
  Meyer}}]{des64}
\bibinfo{author}{\bibfnamefont{J.~A.~D.} \bibnamefont{Shong}},
  \bibinfo{author}{\bibfnamefont{R.~H.} \bibnamefont{Hildebrand}},
  \bibnamefont{and} \bibinfo{author}{\bibfnamefont{P.}~\bibnamefont{Meyer}},
  \bibinfo{journal}{Phys.\ Rev.\ Lett.} \textbf{\bibinfo{volume}{12}},
  \bibinfo{pages}{3} (\bibinfo{year}{1964}).

\bibitem[{\citenamefont{Fanselow et~al.}(1969)\citenamefont{Fanselow, Hartman,
  Hildebrad, and Meyer}}]{fan69}
\bibinfo{author}{\bibfnamefont{J.~L.} \bibnamefont{Fanselow}},
  \bibinfo{author}{\bibfnamefont{R.~C.} \bibnamefont{Hartman}},
  \bibinfo{author}{\bibfnamefont{R.~H.} \bibnamefont{Hildebrad}},
  \bibnamefont{and} \bibinfo{author}{\bibfnamefont{P.}~\bibnamefont{Meyer}},
  \bibinfo{journal}{Astrophys.\ J.} \textbf{\bibinfo{volume}{158}},
  \bibinfo{pages}{771} (\bibinfo{year}{1969}).

\bibitem[{\citenamefont{Adriani et~al.}(2009)}]{adr09b}
\bibinfo{author}{\bibfnamefont{O.}~\bibnamefont{Adriani}} \bibnamefont{et~al.},
  \bibinfo{journal}{Nature} \textbf{\bibinfo{volume}{458}},
  \bibinfo{pages}{607} (\bibinfo{year}{2009}).

\bibitem[{\citenamefont{Delahaye et~al.}(2009)\citenamefont{Delahaye, Donato,
  Fornengo, Lavalle, Lineros, Salati, and Taillet}}]{del09}
\bibinfo{author}{\bibfnamefont{T.}~\bibnamefont{Delahaye}},
  \bibinfo{author}{\bibfnamefont{F.}~\bibnamefont{Donato}},
  \bibinfo{author}{\bibfnamefont{N.}~\bibnamefont{Fornengo}},
  \bibinfo{author}{\bibfnamefont{J.}~\bibnamefont{Lavalle}},
  \bibinfo{author}{\bibfnamefont{R.}~\bibnamefont{Lineros}},
  \bibinfo{author}{\bibfnamefont{P.}~\bibnamefont{Salati}}, \bibnamefont{and}
  \bibinfo{author}{\bibfnamefont{R.}~\bibnamefont{Taillet}},
  \bibinfo{journal}{Astron.\ Astrophys.} \textbf{\bibinfo{volume}{501}},
  \bibinfo{pages}{821} (\bibinfo{year}{2009}).

\bibitem[{\citenamefont{Nishimura et~al.}(1980)}]{nis80}
\bibinfo{author}{\bibfnamefont{J.}~\bibnamefont{Nishimura}}
  \bibnamefont{et~al.}, \bibinfo{journal}{Astrophys.\ J.}
  \textbf{\bibinfo{volume}{238}}, \bibinfo{pages}{394} (\bibinfo{year}{1980}).

\bibitem[{\citenamefont{Delahaye et~al.}(2010)\citenamefont{Delahaye, Lavalle,
  Lineros, Donato, and Fornengo}}]{del10}
\bibinfo{author}{\bibfnamefont{T.}~\bibnamefont{Delahaye}},
  \bibinfo{author}{\bibfnamefont{J.}~\bibnamefont{Lavalle}},
  \bibinfo{author}{\bibfnamefont{R.}~\bibnamefont{Lineros}},
  \bibinfo{author}{\bibfnamefont{F.}~\bibnamefont{Donato}}, \bibnamefont{and}
  \bibinfo{author}{\bibfnamefont{N.}~\bibnamefont{Fornengo}},
  \bibinfo{journal}{Astron.\ Astrophys.} \textbf{\bibinfo{volume}{524}},
  \bibinfo{pages}{A51} (\bibinfo{year}{2010}).

\bibitem[{\citenamefont{Cirelli et~al.}(2008)\citenamefont{Cirelli, Kadastik,
  Raidal, and Strumia}}]{cir08}
\bibinfo{author}{\bibfnamefont{M.}~\bibnamefont{Cirelli}},
  \bibinfo{author}{\bibfnamefont{M.}~\bibnamefont{Kadastik}},
  \bibinfo{author}{\bibfnamefont{M.}~\bibnamefont{Raidal}}, \bibnamefont{and}
  \bibinfo{author}{\bibfnamefont{A.}~\bibnamefont{Strumia}},
  \bibinfo{journal}{Nucl.\ Phys.\ B} \textbf{\bibinfo{volume}{813}},
  \bibinfo{pages}{1} (\bibinfo{year}{2008}).

\bibitem[{\citenamefont{Atoyan et~al.}(1995)\citenamefont{Atoyan, Aharonian,
  and Volk}}]{ato95}
\bibinfo{author}{\bibfnamefont{A.~M.} \bibnamefont{Atoyan}},
  \bibinfo{author}{\bibfnamefont{F.~A.} \bibnamefont{Aharonian}},
  \bibnamefont{and} \bibinfo{author}{\bibfnamefont{H.~J.} \bibnamefont{Volk}},
  \bibinfo{journal}{Phys.\ Rev.\ D} \textbf{\bibinfo{volume}{52}},
  \bibinfo{pages}{3265} (\bibinfo{year}{1995}).

\bibitem[{\citenamefont{Picozza et~al.}(2007)}]{pic07}
\bibinfo{author}{\bibfnamefont{P.}~\bibnamefont{Picozza}} \bibnamefont{et~al.},
  \bibinfo{journal}{Astropart.\ Phys.} \textbf{\bibinfo{volume}{27}},
  \bibinfo{pages}{296} (\bibinfo{year}{2007}).

\bibitem[{\citenamefont{Boezio et~al.}(2009)}]{boe09}
\bibinfo{author}{\bibfnamefont{M.}~\bibnamefont{Boezio}} \bibnamefont{et~al.},
  \bibinfo{journal}{New\ J.\ Phys.} \textbf{\bibinfo{volume}{11}},
  \bibinfo{pages}{105023} (\bibinfo{year}{2009}).

\bibitem[{\citenamefont{Adriani et~al.}(2010{\natexlab{a}})}]{adr10b}
\bibinfo{author}{\bibfnamefont{O.}~\bibnamefont{Adriani}} \bibnamefont{et~al.},
  \bibinfo{journal}{Phys.\ Rev.\ Lett.} \textbf{\bibinfo{volume}{105}},
  \bibinfo{pages}{121101} (\bibinfo{year}{2010}{\natexlab{a}}).

\bibitem[{\citenamefont{Boezio et~al.}(2006)}]{boe06}
\bibinfo{author}{\bibfnamefont{M.}~\bibnamefont{Boezio}} \bibnamefont{et~al.},
  \bibinfo{journal}{Astropart.\ Phys.} \textbf{\bibinfo{volume}{26}},
  \bibinfo{pages}{111} (\bibinfo{year}{2006}).

\bibitem[{\citenamefont{Brun et~al.}(1994)}]{bru94}
\bibinfo{author}{\bibfnamefont{R.}~\bibnamefont{Brun}} \bibnamefont{et~al.},
  \emph{\bibinfo{title}{Detector description and simulation tool}},
  \bibinfo{howpublished}{CERN program library} (\bibinfo{year}{1994}),
  \bibinfo{note}{version 3.21}.

\bibitem[{\citenamefont{Agostinelli et~al.}(2003)}]{ago03}
\bibinfo{author}{\bibfnamefont{S.}~\bibnamefont{Agostinelli}}
  \bibnamefont{et~al.}, \bibinfo{journal}{Nucl.\ Instrum.\ Meth.\ A}
  \textbf{\bibinfo{volume}{506}}, \bibinfo{pages}{250} (\bibinfo{year}{2003}).

\bibitem[{\citenamefont{Adriani et~al.}()}]{adr10c}
\bibinfo{author}{\bibfnamefont{O.}~\bibnamefont{Adriani}} \bibnamefont{et~al.},
  \bibinfo{note}{accepted for pubblication in Science}.

\bibitem[{\citenamefont{Lafferty and Wyatt}(1995)}]{laf95}
\bibinfo{author}{\bibfnamefont{G.~D.} \bibnamefont{Lafferty}} \bibnamefont{and}
  \bibinfo{author}{\bibfnamefont{T.~T.} \bibnamefont{Wyatt}},
  \bibinfo{journal}{Nucl.\ Instrum.\ Meth.\ A} \textbf{\bibinfo{volume}{355}},
  \bibinfo{pages}{541} (\bibinfo{year}{1995}).

\bibitem[{\citenamefont{D'Agostini}(1995)}]{dag95}
\bibinfo{author}{\bibfnamefont{G.}~\bibnamefont{D'Agostini}},
  \bibinfo{journal}{Nucl.\ Instrum.\ Meth.\ A} \textbf{\bibinfo{volume}{362}},
  \bibinfo{pages}{487} (\bibinfo{year}{1995}).

\bibitem[{\citenamefont{Boezio et~al.}(2002)\citenamefont{Boezio, Bonvicini,
  Mocchiutti, Schiavon, Scian, Vacchi, Zampa, and Zampa}}]{boe02}
\bibinfo{author}{\bibfnamefont{M.}~\bibnamefont{Boezio}},
  \bibinfo{author}{\bibfnamefont{V.}~\bibnamefont{Bonvicini}},
  \bibinfo{author}{\bibfnamefont{E.}~\bibnamefont{Mocchiutti}},
  \bibinfo{author}{\bibfnamefont{P.}~\bibnamefont{Schiavon}},
  \bibinfo{author}{\bibfnamefont{G.}~\bibnamefont{Scian}},
  \bibinfo{author}{\bibfnamefont{A.}~\bibnamefont{Vacchi}},
  \bibinfo{author}{\bibfnamefont{G.}~\bibnamefont{Zampa}}, \bibnamefont{and}
  \bibinfo{author}{\bibfnamefont{N.}~\bibnamefont{Zampa}},
  \bibinfo{journal}{Nucl.\ Instrum.\ Meth.\ A} \textbf{\bibinfo{volume}{487}},
  \bibinfo{pages}{407} (\bibinfo{year}{2002}).

\bibitem[{\citenamefont{Boezio et~al.}(2000)}]{boe00}
\bibinfo{author}{\bibfnamefont{M.}~\bibnamefont{Boezio}} \bibnamefont{et~al.},
  \bibinfo{journal}{Astrophys.\ J.} \textbf{\bibinfo{volume}{532}},
  \bibinfo{pages}{653} (\bibinfo{year}{2000}).

\bibitem[{\citenamefont{DuVernois et~al.}(2001)}]{duv01}
\bibinfo{author}{\bibfnamefont{M.~A.} \bibnamefont{DuVernois}}
  \bibnamefont{et~al.}, \bibinfo{journal}{Astrophys.\ J.}
  \textbf{\bibinfo{volume}{559}}, \bibinfo{pages}{296} (\bibinfo{year}{2001}).

\bibitem[{\citenamefont{Alcaraz et~al.}(2000)}]{alc00}
\bibinfo{author}{\bibfnamefont{J.}~\bibnamefont{Alcaraz}} \bibnamefont{et~al.},
  \bibinfo{journal}{Phys.\ Lett.\ B} \textbf{\bibinfo{volume}{484}},
  \bibinfo{pages}{10} (\bibinfo{year}{2000}).

\bibitem[{\citenamefont{Grimani et~al.}(2002)}]{gri02}
\bibinfo{author}{\bibfnamefont{C.}~\bibnamefont{Grimani}} \bibnamefont{et~al.},
  \bibinfo{journal}{Astron.\ Astrophys.} \textbf{\bibinfo{volume}{392}},
  \bibinfo{pages}{287} (\bibinfo{year}{2002}).

\bibitem[{\citenamefont{Kobayashi et~al.}(1999)\citenamefont{Kobayashi,
  Nishimura, Komori, Shirai, Tateyama, Taira, Yoshida, and Yuda}}]{kob99}
\bibinfo{author}{\bibfnamefont{T.}~\bibnamefont{Kobayashi}},
  \bibinfo{author}{\bibfnamefont{J.}~\bibnamefont{Nishimura}},
  \bibinfo{author}{\bibfnamefont{Y.}~\bibnamefont{Komori}},
  \bibinfo{author}{\bibfnamefont{T.}~\bibnamefont{Shirai}},
  \bibinfo{author}{\bibfnamefont{N.}~\bibnamefont{Tateyama}},
  \bibinfo{author}{\bibfnamefont{T.}~\bibnamefont{Taira}},
  \bibinfo{author}{\bibfnamefont{K.}~\bibnamefont{Yoshida}}, \bibnamefont{and}
  \bibinfo{author}{\bibfnamefont{T.}~\bibnamefont{Yuda}}, in
  \emph{\bibinfo{booktitle}{Proc. 26th Int. Cosmic Ray Conf. (Salt Lake City)}}
  (\bibinfo{year}{1999}), vol.~\bibinfo{volume}{3}, p.~\bibinfo{pages}{61}.

\bibitem[{\citenamefont{Torii et~al.}(2001)}]{tor01}
\bibinfo{author}{\bibfnamefont{S.}~\bibnamefont{Torii}} \bibnamefont{et~al.},
  \bibinfo{journal}{Astrophys.\ J.} \textbf{\bibinfo{volume}{559}},
  \bibinfo{pages}{973} (\bibinfo{year}{2001}).

\bibitem[{\citenamefont{Chang et~al.}(2008)}]{cha08}
\bibinfo{author}{\bibfnamefont{J.}~\bibnamefont{Chang}} \bibnamefont{et~al.},
  \bibinfo{journal}{Nature} \textbf{\bibinfo{volume}{456}},
  \bibinfo{pages}{362} (\bibinfo{year}{2008}).

\bibitem[{\citenamefont{Aharonian et~al.}(2008)}]{aha08}
\bibinfo{author}{\bibfnamefont{F.}~\bibnamefont{Aharonian}}
  \bibnamefont{et~al.}, \bibinfo{journal}{Phys.\ Rev.\ Lett.}
  \textbf{\bibinfo{volume}{101}}, \bibinfo{pages}{261104}
  (\bibinfo{year}{2008}).

\bibitem[{\citenamefont{Ackermann et~al.}(2010)}]{ack10}
\bibinfo{author}{\bibfnamefont{M.}~\bibnamefont{Ackermann}}
  \bibnamefont{et~al.}, \bibinfo{journal}{Phys.\ Rev.\ D}
  \textbf{\bibinfo{volume}{82}}, \bibinfo{pages}{092004}
  (\bibinfo{year}{2010}).

\bibitem[{\citenamefont{Adriani et~al.}(2010{\natexlab{b}})}]{adr10a}
\bibinfo{author}{\bibfnamefont{O.}~\bibnamefont{Adriani}} \bibnamefont{et~al.},
  \bibinfo{journal}{Astropart.\ Phys.} \textbf{\bibinfo{volume}{34}},
  \bibinfo{pages}{1} (\bibinfo{year}{2010}{\natexlab{b}}).

\bibitem[{\citenamefont{Strong and Moskalenko}(1998)}]{str98}
\bibinfo{author}{\bibfnamefont{A.~W.} \bibnamefont{Strong}} \bibnamefont{and}
  \bibinfo{author}{\bibfnamefont{I.~V.} \bibnamefont{Moskalenko}},
  \bibinfo{journal}{Astrophys.\ J.} \textbf{\bibinfo{volume}{509}},
  \bibinfo{pages}{212} (\bibinfo{year}{1998}).

\bibitem[{\citenamefont{Ptuskin et~al.}(2006)}]{ptu06}
\bibinfo{author}{\bibfnamefont{V.~S.} \bibnamefont{Ptuskin}}
  \bibnamefont{et~al.}, \bibinfo{journal}{Astrophys.\ J.}
  \textbf{\bibinfo{volume}{642}}, \bibinfo{pages}{902} (\bibinfo{year}{2006}).

\bibitem[{\citenamefont{Gleeson and Axford}(1968)}]{gle68}
\bibinfo{author}{\bibfnamefont{L.~J.} \bibnamefont{Gleeson}} \bibnamefont{and}
  \bibinfo{author}{\bibfnamefont{W.~I.} \bibnamefont{Axford}},
  \bibinfo{journal}{Astrophys.\ J.} \textbf{\bibinfo{volume}{154}},
  \bibinfo{pages}{1011} (\bibinfo{year}{1968}).

\bibitem[{\citenamefont{Shaviv et~al.}(2009)\citenamefont{Shaviv, Nakar, and
  Piran}}]{sha09}
\bibinfo{author}{\bibfnamefont{N.~J.} \bibnamefont{Shaviv}},
  \bibinfo{author}{\bibfnamefont{E.}~\bibnamefont{Nakar}}, \bibnamefont{and}
  \bibinfo{author}{\bibfnamefont{T.}~\bibnamefont{Piran}},
  \bibinfo{journal}{Phys.\ Rev.\ Lett.} \textbf{\bibinfo{volume}{103}},
  \bibinfo{pages}{111302} (\bibinfo{year}{2009}).

\end{thebibliography}

\end{document}